%% file: main_new.tex
\documentclass{article}
\usepackage{spconf,amsmath,graphicx,hyperref}
\usepackage{cite}
\usepackage{amsfonts}
\usepackage{algorithmic}
\usepackage{graphicx}
\usepackage{textcomp}
\usepackage{xcolor}
\usepackage{booktabs}
\usepackage{float}  

\title{Retaining Mixture Representations for Domain Generalized \\ Anomalous Sound Detection}

\name{Phurich Saengthong$^{1,2,\dagger}$\thanks{${\dagger }$Work done during internship at Hitachi.}, Tomoya Nishida$^2$, Kota Dohi$^2$,
Natsuo Yamashita$^2$, Yohei Kawaguchi$^2$}

\address{$^1$Institute of Science Tokyo,
         $^2$R\&D Group, Hitachi Ltd., Japan}

\begin{document}
\ninept
\maketitle
\begin{abstract}
Anomalous sound detection (ASD) in the wild requires robustness to distribution shifts such as unseen low-SNR input mixtures of machine and noise types. State-of-the-art systems extract embeddings from an adapted audio encoder and detect anomalies via nearest-neighbor search, but fine-tuning on noisy machine sounds often acts like a denoising objective, suppressing noise and reducing generalization under mismatched mixtures or inconsistent labeling. Training-free systems with frozen self-supervised learning (SSL) encoders avoid this issue and show strong first-shot generalization, yet their performance drops when mixture embeddings deviate from clean-source embeddings. We propose to improve SSL backbones with a \textit{retain-not-denoise} strategy that better preserves information from mixed sound sources. The approach combines a multi-label audio tagging loss with a mixture alignment loss that aligns student mixture embeddings to convex teacher embeddings of clean and noise inputs. Controlled experiments on stationary, non-stationary, and mismatched noise subsets demonstrate improved robustness under distribution shifts, narrowing the gap toward oracle mixture representations.
\end{abstract}
\vspace{-1mm}
\section{Introduction}

\input{introduction}

\begin{figure}[!h]
    \centering
    \includegraphics[width=0.92\columnwidth]{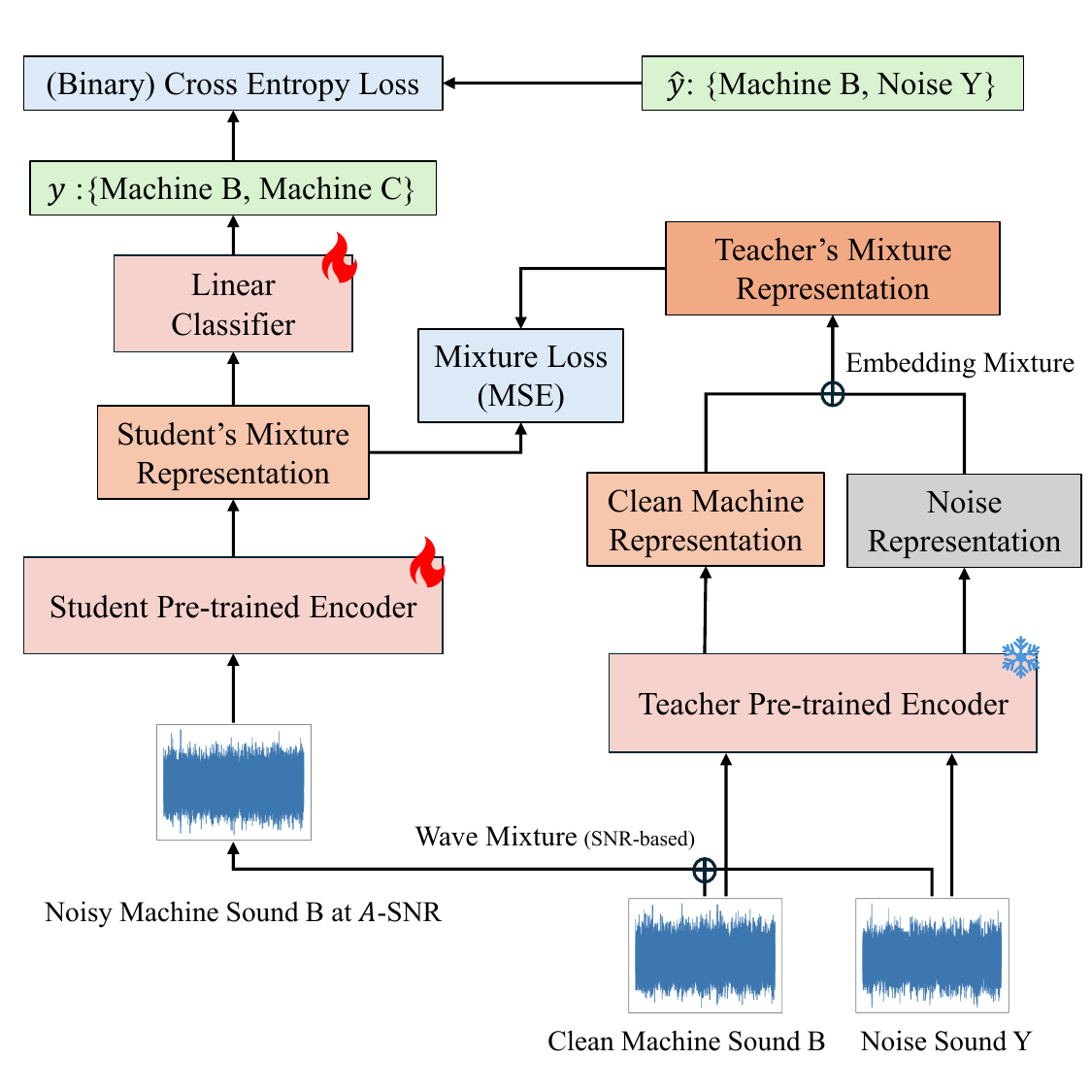}
    \vspace{-3mm}
    \caption{Overview of the proposed approach.}
    \label{fig:proposed_method}
     \vspace{-5mm}
\end{figure}

\section{Audio Encoder for ASD}

\input{related_work}

\section{Pretraining Audio Encoder via Retaining Mixture Representation}
\input{method}

\vspace{-2mm}
\section{Experiments}
\input{experiments}

\section{Conclusion}
We proposed a \textit{retain-not-denoise} pre-training strategy that combines tagging and mixture losses to improve the robustness of SSL-based audio encoders for ASD. Through controlled experiments, we showed that training on machine data, even when drawn from distributions disjoint from evaluation, effectively builds training-free embedding extractors under domain shifts. Crucially, by retaining information from both machine and noise sources, the approach provides more reliable representations of mixed audio and improves upon SSL backbone baselines, unlike recent denoising methods that tend to degrade performance under low-SNR conditions. These results highlight the importance of preserving full input mixture information and demonstrate that machine data from different distributions can still strengthen training, pointing toward more generalizable feature extractors and the development of a universal ASD system.


\bibliographystyle{IEEEbib}
\clearpage
\footnotesize
\bibliography{bib}

\end{document}

%% file: introduction.tex
Anomalous sound detection (ASD) is a key approach for monitoring machine condition and ensuring reliable operation. A foundational goal is to build systems that work reliably across different machine types and operating conditions, even under distribution shifts such as changing background noise or environments. To address this challenge, recent research has increasingly focused on developing more robust embedding extractors. \cite{koizumi_description_2020, dohi_description_2023, Nishida_arXiv2024_01, Nishida_arXiv2025_01}


Over the past few years, discriminative approaches with machine labels have shown strong performance \cite{Giri2020, wilkinghoff_sub-cluster_2021, liu_anomalous_2022full, wilkinghoff_self-supervised_2023_full, wilkinghoff_why_2024, Jiang2024AnoPatchTB, 10890020}. These methods classify both target and non-target machine classes, constraining embeddings within machine-specific boundaries and leveraging outlier exposure \cite{koizumi_description_2020, hendrycks2019oe, Giri2020}. More recently, fine-tuning self-supervised encoders pretrained on large-scale audio data has yielded embeddings more robust to domain shifts, which is particularly effective for ASD where long-tail machine conditions are common \cite{Jiang2024AnoPatchTB, fujimura2025asdkit}.

However, recent discriminative approaches often rely on target machine sounds for training or fine-tuning, making them less suitable than training-free systems, where frozen SSL encoders are directly paired with reference sounds and nearest-neighbor search \cite{genrep}. The latter can operate out-of-the-box without fine-tuning on target references. Moreover, when encoders are trained in a discriminative manner on noisy mixtures with only machine attribution labels, they learn to separate machine classes while suppressing noise \cite{koizumi_description_2020, dohi_description_2023, Nishida_arXiv2024_01, Nishida_arXiv2025_01}. This implicit denoising \cite{niizumi2024m2dx} can hinder generalization: unseen noise may resemble machine sounds under low SNR, or denoising strategies in training may mismatch those in evaluation. These limitations were evident in the DCASE2025 Task 2 first-shot evaluation \cite{Nishida_arXiv2025_01}, where large fine-tuned SSL systems \cite{JiangTHUEE2025, ZhengSJTU-AITHU2025} trained with additional DCASE datasets (e.g., DCASE2020T2) underperformed compared to training-free methods \cite{genrep_firstshot}. A key issue is that the additional datasets introduced different input mixtures and machine label definitions, sometimes involving the same machine type but with different labeling criteria, creating a mismatch between pre-training and evaluation tasks and causing fine-tuning to degrade performance relative to the unadapted backbone.

Although SSL backbones themselves have shown strong performance for ASD tasks as training-free feature extractors \cite{genrep, wu2025towards, genrep_firstshot, Nishida_arXiv2025_01}, their representations remain susceptible to noise. In particular, under low SNR conditions they struggle to capture the full information of mixture inputs, limiting robustness in practical deployments. Furthermore, it remains unclear under which conditions, for example different noise types or target SNRs, current methods achieve their gains. 

To address this gap, we design controlled experiments with explicit evaluation settings to isolate key factors and gain clearer insights into robustness, which in turn motivates our proposed approach. We explore how to improve off-the-shelf SSL backbones as training-free embedding extractors, aiming for direct application to unseen input mixtures of machine and noise types toward a general ASD system. Our analysis reveals a fundamental limitation: the backbone does not represent mixtures of machine and noise sounds effectively, whereas averaging the embeddings of each source extracted separately produces stronger representations (Table \ref{tab:feature_alignment}). We regard this combined embedding as an oracle representation, exposing a mismatch between oracle and mixture features. This motivates our feature alignment strategy. Building on this insight, we propose a pre-training method based on a \textit{retain-not-denoise} strategy, combining (i) an audio tagging loss that classifies both machine and noise events, and (ii) a mixture loss that aligns student encoder outputs of noisy mixtures with oracle embeddings from a frozen teacher backbone. In controlled SNR-based experiments, our method improves upon the BEATs iter3 \cite{pmlr-v202-chen23ag} backbone and outperforms discriminative-based approaches that implicitly rely on denoising.
\vspace{-3mm}



%% file: related_work.tex
\noindent \textbf{Discriminative Learning for Target Audio Encoder.}
Discriminative learning trains or fine-tunes audio encoders to classify machine types using cross-entropy loss. Given a mixed sample $\mathbf{x}_{\text{mix}}$, only the machine label is used, with the noise component ignored \cite{Giri2020, wilkinghoff_sub-cluster_2021, liu_anomalous_2022full, wilkinghoff_self-supervised_2023_full, wilkinghoff_why_2024, Jiang2024AnoPatchTB, 10890020}. A one-hot vector $\hat{y} \in \{0,1\}^C$ indicates the target machine class, and the encoder output $y = f(\mathbf{x}_{\text{mix}})$ is optimized with binary cross-entropy:
\begin{equation}
    \mathcal{L}_{\text{denoise}} 
    = - \sum_{c=1}^{C} \big[ \hat{y}_c \log y_c + (1-\hat{y}_c) \log (1-y_c) \big].
\end{equation}

To improve embedding separation, angular margin losses are often applied \cite{liu_anomalous_2022full, wilkinghoff_sub-cluster_2021, wilkinghoff_self-supervised_2023_full, Jiang2024AnoPatchTB}, and linear mixup is used to enhance robustness through pseudo anomalies \cite{zhang2018mixup, wilkinghoff_sub-cluster_2021}.

\noindent \textbf{ASD using Pre-trained Audio Foundation Models.}
Recent studies show that SSL audio foundation models \cite{audioset, pmlr-v202-chen23ag, gong21b_interspeech, ijcai2024p421, alex2025sslam, niizumi2024m2d-clap} can serve as training-free feature extractors for ASD \cite{genrep, wu2025towards}. Given a test input $X_{\text{test}}$, the encoder $\Phi(\cdot)$ produces frame-level embeddings $f \in \mathbb{R}^{L \times D}$. Following \cite{genrep}, we reshape $L = T \times F$ into $f \in \mathbb{R}^{T \times F \times D}$, apply mean pooling along $T$, and flatten to $f \in \mathbb{R}^{F \cdot D}$. Anomaly scores are then computed by KNN distance between the test embedding and reference embeddings:
\begin{equation}
A(X_{\text{test}}) =
\frac{1}{K} \sum_{\mathbf{f} \in \mathcal{N}{K}(\Phi(X{\text{test}});,\Phi(X_{\text{ref}}))}
d(\mathbf{f}, \Phi(X_{\text{test}})),
\end{equation}
where $\mathcal{N}_{K}$ are the $K$ nearest reference embeddings and $d(\cdot)$ is a distance function (e.g., Euclidean).

While effective under clean conditions, SSL encoders can fail when target machine sounds are overlapped by background noise. Under low-SNR mixtures, embeddings lose discriminative information because pre-training objectives (e.g., masked prediction \cite{pmlr-v202-chen23ag} or feature reconstruction \cite{niizumi2024m2dx, ijcai2024p421}) are not designed for mixtures. To mitigate this, SSLAM \cite{alex2025sslam} introduced a mixing-based pre-training on EAT \cite{ijcai2024p421}, which showed gains on audio tagging benchmarks. However, the method represents mixtures by reconstructing only an averaged embedding of two sound sources, computed from the maximum energy of their mixed spectrograms. This averaging ignores the actual power and amplitude relationships between sources, making the strategy unrealistic for real-world mixtures. A more principled approach is to mix signals by target SNRs, ensuring embeddings preserve information from both sources.

%% file: method.tex

We propose a training-free feature extractor that improves robustness to distribution shifts by further pre-training an SSL audio encoder with a \textit{retain-not-denoise} strategy. In contrast to prior work that frames pre-training as an implicit denoising task, we argue that the objective should preserve information from all mixed sound sources. We show that simply retaining noise, rather than suppressing it, leads to better generalization to unseen input mixture distributions. As illustrated in Fig.~\ref{fig:proposed_method}, our approach combines two components: (i) an audio tagging loss that encourages the student encoder to retain both machine and noise information, and (ii) a mixture alignment loss that aligns the student’s mixture representation with oracle embeddings generated by a frozen teacher encoder.

\noindent \textbf{Mixing Audios.} We adopt an SNR-based mixing strategy that normalizes the noise to unit power and adjusts the signal amplitude relative to this baseline. Let $P_1$ and $P_2$ denote the powers of $\mathbf{x}_1$ and $\mathbf{x}_2$, and enforce $(a_1^2 P_1)/(a_2^2 P_2) = R$, where $R = 10^{\mathrm{SNR}_{\mathrm{dB}}/10}$. Setting $a_2^2 P_2 = 1$ yields $a_1 = \sqrt{R/P_1}$ and $a_2 = \sqrt{1/P_2}$. The resulting mixture is:
\begin{equation}
    \mathbf{x}_{\text{mix}} = a_1 \mathbf{x}_1 + a_2 \mathbf{x}_2,
\end{equation}
ensuring the amplitude–power relationship respects the specified SNR.

\noindent \textbf{Classification with Audio Tagging.} Each mixture $\mathbf{x}_{\text{mix}}$ is assigned a multi-hot label vector $\hat{y} \in \{0,1\}^{C+N}$, where $C$ is the number of machine classes and $N$ is the number of noise categories. The student encoder extracts an embedding $f_{\text{student}} = \Phi_{\text{student}}(\mathbf{x}_{\text{mix}})$, which is passed through a linear layer $y = \text{Linear}(f_{\text{student}})$. The objective is the binary cross-entropy:
\begin{equation}
    \mathcal{L}_{\text{tagging}} 
    = - \sum_{k=1}^{C+N} \big[ \hat{y}_k \log y_k + (1-\hat{y}_k) \log (1-y_k) \big].
\end{equation}

\noindent \textbf{Mixture Alignment Loss.}  
In addition to classification, we introduce a feature alignment objective to guide the student encoder. The teacher encoder is a frozen copy of the same SSL backbone (e.g., BEATs \cite{pmlr-v202-chen23ag}), ensuring stable representations without parameter updates \cite{bolya2025perceptionencoderbestvisual}. For each mixture input, the teacher encodes the clean machine and noise signals separately, yielding $\Phi_{\text{teacher}}(\mathbf{x}_{\text{target}}), \Phi_{\text{teacher}}(\mathbf{x}_{\text{noise}}) \in \mathbb{R}^{L \times D}$, where $L (T \times F)$ is the sequence length and $D$ the feature dimension. A mixture-consistent target is then obtained by an element-wise convex combination:
\begin{equation}
    f_{\text{teacher}} = \lambda \,\Phi_{\text{teacher}}(\mathbf{x}_{\text{target}}) 
                       + (1-\lambda)\,\Phi_{\text{teacher}}(\mathbf{x}_{\text{noise}}).
\end{equation}

The student encoder, which receives only the raw mixture $\mathbf{x}_{\text{mix}}$, produces an embedding $f_{\text{student}} = \Phi_{\text{student}}(\mathbf{x}_{\text{mix}})$. We align the student’s mixture embedding to the teacher’s mixture-consistent target using mean squared error, similar to \cite{alex2025sslam}:
\begin{equation}
    \mathcal{L}_{\text{mixture}} = \lVert f_{\text{student}} - f_{\text{teacher}} \rVert_2^2.
\end{equation}
In this work, we use a fixed $\lambda = 0.5$ in the embedding space for stability; adapting $\lambda$ to reflect input SNR remains an interesting direction for future work.

\noindent \textbf{Overall Objective.} 
The final training objective combines classification and alignment:
\begin{equation}
    \mathcal{L} = \alpha \,\mathcal{L}_{\text{tagging}} + \beta \,\mathcal{L}_{\text{mixture}},
\end{equation}
where $\alpha$ and $\beta$ control the relative importance of retaining class information versus aligning mixture embeddings. 

Together, these objectives implement the \textit{retain-not-denoise} strategy, encouraging the encoder to preserve machine–noise mixtures while remaining robust under distribution shifts. Our method differs from prior approaches in two ways: (i) unlike classification-based objectives that suppress noise, it retains both machine and noise information \cite{koizumi_description_2020, niizumi2024m2dx}, and (ii) unlike \cite{alex2025sslam, mixit}, it aligns mixture embeddings to convex teacher targets while keeping the teacher frozen for stable supervision. We further investigate pre-training of training-free extractors for ASD, showing that alignment improves robustness. Conceptually, our approach resembles teacher-guided alignment in vision \cite{bolya2025perceptionencoderbestvisual}, which also enhanced downstream performance.

%% file: experiments.tex

\begin{table}[!t]\centering
\caption{Dataset statistics.}
\setlength{\tabcolsep}{3pt}
\label{tab:dataset_stats}
\resizebox{0.93\columnwidth}{!}{%
\begin{tabular}{lcccc}
\toprule
 & \multicolumn{2}{c}{Clean machine sounds} & \multicolumn{2}{c}{Noise sounds} \\
\cmidrule(lr){2-3}\cmidrule(lr){4-5}
 & Normal & Abnormal & Stationary & Non-stationary \\
\midrule
\textbf{Pre-training data} \\
Audios          & 36,411 & -- & 1,537 & 3,122 \\
Machine types  & \multicolumn{2}{c}{8}   & \multicolumn{2}{c}{--} \\
Attribute classes  & \multicolumn{2}{c}{231} & \multicolumn{2}{c}{4} \\
\midrule
\textbf{Evaluation data (Per machine type)} \\
Reference (Normal) & \multicolumn{4}{c}{990 source / 10 target} \\
Testing (Normal/Abnormal) & \multicolumn{4}{c}{50 source + 50 target each} \\
\bottomrule
\vspace{-10mm}
\end{tabular}
}
\end{table}
\vspace{-3mm}

\input{experiments/result_main}
\subsection{Setup}
\input{setup}

\input{experiments/finetuning_foundation}  

%% file: experiments/result_main.tex
\begin{table*}[!th]\centering
\caption{Performance comparison of baseline and proposed encoder methods under different noise conditions (higher is better).}\label{tab:main_results}
\setlength{\tabcolsep}{3.0 pt}
\resizebox{0.93\textwidth}{!}{%
\begin{tabular}{l|ccccccc|ccccccc|ccccccc|cc}
\toprule
&\multicolumn{7}{c|}{Factory A (Stationary)} 
 &\multicolumn{7}{c|}{Factory B (Non-stationary)} 
 &\multicolumn{7}{c|}{Mismatch (Factory B $\longrightarrow$ Factory A)} 
 &\multicolumn{2}{c}{Hmean}\\\cmidrule{2-8}\cmidrule{9-15}\cmidrule{16-22}
Audio Encoder \cite{genrep} &-10 &-5 &0 &5 &10 &20 &30 &-10 &-5 &0 &5 &10 &20 &30 &-10 &-5 &0 &5 &10 &20 &30 &$\{-10,\,-5,\,0\}$ &All \\\midrule
EAT-10 \cite{ijcai2024p421} &60.0 &73.6 &85.6 &89.3 &92.1 &94.5 &94.9 &68.3 &76.2 &84.9 &90.2 &93.2 &94.7 &94.9 &46.5 &47.2 &57.0 &70.3 &80.5 &88.4 &91.1 &63.5 &75.9 \\
EAT-30 \cite{ijcai2024p421} &61.3 &74.2 &85.6 &90.5 &92.3 &95.2 &95.8 &68.4 &76.9 &86.8 &92.4 &94.3 &95.5 &96.3 &46.7 &51.9 &61.8 &71.8 &81.1 &90.1 &92.7 &65.5 &77.6 \\
SSLAM \cite{alex2025sslam} &59.3 &71.7 &84.7 &89.7 &91.6 &94.6 &94.7 &65.5 &74.9 &83.8 &90.6 &93.1 &94.2 &94.6 &47.5 &47.6 &60.6 &72.7 &82.3 &89.1 &91.5 &63.5 &76.1 \\
BEATs iter3 \cite{pmlr-v202-chen23ag} &62.0 &\textbf{77.2} &86.8 &91.2 &93.9 &96.5 &97.3 &69.3 &78.3 &86.3 &92.1 &95.1 &96.6 &97.3 &47.7 &50.9 &61.5 &73.8 &83.5 &91.5 &94.4 &66.0 &78.5 \\
\midrule
Denoising baseline &64.3 &74.6 &87.0 &90.6 &92.2 &94.1 &94.9 &67.0 &75.0 &84.7 &90.9 &93.1 &94.1 &94.8 &45.2 &49.8 &61.3 &72.4 &83.3 &90.7 &92.6 &64.7 &77.0 \\
\hspace{0.5em}+ Linear mixup \cite{zhang2018mixup} (mixture of mixture) &62.8 &74.9 &83.0 &88.3 &90.6 &92.8 &94.2 &67.0 &74.5 &84.3 &89.9 &92.1 &93.2 &93.8 &45.2 &45.3 &55.7 &67.8 &78.5 &88.4 &92.5 &62.6 &74.9 \\
\hspace{0.5em}+ SNR mixup (mixture of mixture) &\textbf{64.4} &76.4 &85.0 &90.3 &92.3 &94.9 &95.7 &68.2 &77.1 &85.4 &92.0 &94.3 &95.4 &96.0 &48.1 &49.0 &59.6 &72.0 &82.2 &89.8 &93.4 &65.4 &77.5 \\
\midrule
Ours - Mixing Eq.~(3) at $\pm 5$ dB & & & & & & & & & & & & & & & & & & & & & & & \\
\textit{Tagging Loss} ($\alpha=1, \beta=0$) &63.2 &76.0 &86.4 &91.3 &94.3 &97.2 &97.6 &71.2 &80.3 &87.4 &92.5 &95.4 &97.1 &97.5 &49.1 &55.9 &62.8 &71.2 &81.0 &91.1 &94.6 &67.8 &79.4 \\
\textit{Mixture Loss} ($\alpha=0, \beta=1$) &63.9 &76.5 &\textbf{88.7} &92.9 &95.1 &97.2 &97.9 &71.4 &80.6 &89.1 &94.2 &96.3 &97.1 &97.8 &50.7 &56.6 &64.6 &74.6 &84.4 &92.9 &95.7 &69.0 &80.7 \\
\textit{Tagging Loss} + \textit{Mixture Loss} ($\alpha=1, \beta=1$)&63.5 &76.6 &88.4 &92.9 &94.7 &96.8 &97.5 &71.2 &80.5 &88.9 &94.4 &96.1 &97.0 &97.5 &50.0 &55.9 &64.5 &75.0 &84.2 &92.8 &95.5 &68.6 &80.4 \\
\midrule
Ours - Mixing Eq.~(3) at $0$ dB & & &\textbf{} &\textbf{} &\textbf{} & & & & & &\textbf{} &\textbf{} &\textbf{} & & & &\textbf{} &\textbf{} &\textbf{} &\textbf{} &\textbf{} & &\textbf{} \\
\textit{Tagging Loss} ($\alpha=1, \beta=0$) &63.6 &75.6 &87.6 &\textbf{93.7} &\textbf{95.9} &\textbf{97.9} &\textbf{98.2} &69.8 &80.6 &\textbf{90.2} &\textbf{95.5} &\textbf{97.0} &\textbf{98.1} &\textbf{98.2} &\textbf{53.3} &57.8 &65.6 &\textbf{77.9} &\textbf{87.0} &\textbf{95.4} &\textbf{97.2} &69.5 &\textbf{81.6} \\
\textit{Mixture Loss} ($\alpha=0, \beta=1$) &63.0 &76.9 &88.3 &93.1 &95.2 &97.3 &97.8 &\textbf{71.8} &\textbf{81.5} &89.8 &94.7 &96.3 &97.1 &97.7 &51.7 &58.7 &67.8 &76.2 &84.6 &92.6 &95.6 &70.0 &81.4 \\
\textit{Tagging Loss} + \textit{Mixture Loss} ($\alpha=1, \beta=1$) &62.7 &76.5 &88.4 &92.8 &95.1 &97.8 &\textbf{98.2} &71.7 &\textbf{81.5} &89.7 &94.7 &96.8 &97.6 &98.1 &52.5 &\textbf{59.2} &\textbf{68.2} &76.5 &84.6 &93.6 &96.6 &\textbf{70.1} &\textbf{81.6} \\
\bottomrule
\end{tabular}
}
\vspace{-3mm}
\end{table*}

%% file: setup.tex
As shown in Table~\ref{tab:dataset_stats}, we construct two non-overlapping corpora (both machine and noise classes) for pre-training and evaluation of training-free encoders under distribution shifts.

The pre-training set contains eight machine types: bearing, gearbox, fan, slider, and valve from MIMII DG \cite{dohi_mimii_2022}, together with 3DPrinter, AirCompressor, and Scanner from ToyADMOS+ \cite{haradatoyadmos2+}. It also includes four noise attribution classes derived from real factory recordings.

The evaluation set uses six disjoint machine types: \textit{bandsaw}, \textit{grinder}, \textit{shaker}, \textit{screwfeeder}, \textit{bandsealer}, and \textit{polisher} from ToyADMOS+ \cite{haradatoyadmos2+}. To assess robustness, we design three controlled subsets: (i) \textbf{Factory A}, mixing clean machine audio with stationary factory noise; (ii) \textbf{Factory B}, mixing with non-stationary noise; and (iii) \textbf{Mismatch}, where reference clips use non-stationary noise while test clips use stationary noise. This setup creates explicit distribution shifts between reference and test samples.

For noise control, we set target SNRs separately for each machine type in the range $-10$ to $30$ dB. Each noise waveform is scaled relative to the average power of the clean signal before mixing, ensuring that the expected SNR matches the target. Noise clips are used without replacement so that every clean machine recording is paired with a unique noise segment. This procedure standardizes SNRs at the dataset level while preserving instance-level variation to reflect realistic operating conditions.

All recordings are 10 s at 16 kHz, and split-wise counts (source/target, reference/testing) are summarized in Table~\ref{tab:dataset_stats}. For completeness, we also report additional results on DCASE2023T2 \cite{dohi_description_2023} and DCASE2025T2 \cite{Nishida_arXiv2025_01}, whose machine and noise data are excluded from pre-training.

\noindent{\textbf{Metrics.}} For each test subset, we report the official score from the DCASE Task 2 challenge \cite{dohi_description_2023}, defined as the harmonic mean of Source AUC, Target AUC, and pAUC computed over all test data (source and target combined).

\vspace{-3mm}
\subsection{Implementation Details}
All audio is resampled to 16 kHz and constrained to 10 s by truncation or zero-padding. Input features are 128-dim Mel-filterbanks (25 ms Povey window, 10 ms frame shift), normalized with AudioSet statistics. For the \textbf{SSL encoder baselines} (BEATs, EAT, SSLAM \cite{pmlr-v202-chen23ag, ijcai2024p421, alex2025sslam}), we use the default feature pipeline and directly extract representations. For the \textbf{denoising baseline}, we use BEATs iter3 fineunted on 231 machine classes (Eq.~(1)) with randomly mixed noise (Eq.~(3)). On top of this baseline, we consider two extensions: (1) applying linear mixup \cite{zhang2018mixup}, where sampling ratios are drawn from a Beta distribution, and (2) applying SNR mixup (Eq.~(3)) with hard labels. For the \textbf{proposed method}, BEATs iter3 finetuned on 235 classes (machine + noise). Noise is mixed into training inputs, and the task is formulated as audio tagging. A mixture alignment loss ($\lambda=0.5$) is added, and the same SpecAugment is applied only to the input of mixture. Both denoising and proposed models are trained for 20k steps (batch 64, grad accumulation 2) with the AdamW optimizer (learning rate $1\times10^{-4}$, weight decay $1\times10^{-3}$) and a cosine scheduler with 200 warmup steps. Weighted sampling is used, to addresses class imbalance, and unless noted, SNRs are uniformly sampled from $-5$ to $5$\,dB. For \textbf{evaluation}, we follow \cite{genrep} and apply $k$-nearest neighbor classification with $k=1$. For SSL backbones, representations are taken from the final (12th) transformer layer. For the denoising and retraining pre-training, we instead use representations from the 6th layer. This choice is motivated by prior work \cite{genrep} and further supported by our preliminary results, which also show that when models are trained on data from domains different from the target machines, intermediate layers yield more robust representations than the final layer.

%% file: experiments/finetuning_foundation.tex
 



\vspace{-3mm}
\subsection{Main Results}  
Table~\ref{tab:main_results} compares SSL backbones, denoising baselines, and our proposed method under the GenRep framework \cite{genrep}. Among SSL backbones, BEATs achieved the best overall performance; EAT-30 (30 epochs) outperformed both EAT-10 (10 epochs) and SSLAM, indicating that SSLAM’s mixture pre-training objective is less effective for ASD robustness benchmarks. Denoising pre-training further degraded performance compared to BEATs iter3, consistent with recent DCASE2025 Task 2 findings \cite{Nishida_arXiv2025_01, JiangTHUEE2025, ZhengSJTU-AITHU2025, genrep}. In contrast, our proposed method with \textit{Tagging Loss} improved over BEATs by leveraging machine and noise labels, and adding the \textit{Mixture Loss} yielded the best overall performance. While \textit{Tagging Loss} is particularly strong when training with fixed 0 dB mixtures, \textit{Mixture Loss} consistently improved performance in the low-SNR ranges (–10 to 0 dB), both under fixed 0 dB training and when sampling mixtures at $\pm 5$ dB. Combining both objectives provided a more balanced improvement across SNR conditions, with gains of +4.1 over BEATs at the low SNRs and +3.1 on average across all ranges.

\vspace{-3mm}
\input{experiments/result_feature_alignment}

\subsection{Ablation and Analysis}

Table~\ref{tab:feature_alignment} examines feature alignment under the 0 dB SNR condition. As an upper bound, we evaluate embedding mixing, where clean and noise embeddings from the same Mismatch subset are combined using Equation (5) with $\lambda=0.5$. This achieves a large gain over wave mixing (73.6 vs.\ 60.4), showing that averaging embeddings provides a strong target. Our proposed \textit{Mixture Loss} improves upon the baseline (64.2 vs.\ 60.4), though it still falls short of the upper bound (64.2 vs.\ 73.6). These results suggest that while feature alignment helps, there remains significant room to enhance SSL backbone robustness or to design more inherently robust SSL methods.

\input{experiments/result_snr_abla}
Table~\ref{tab:result_snr_ablation} compares the effects of different SNR settings used for pre-training with the \textit{Tagging Loss}. The best performance is achieved when the target SNR for mixing is fixed at 0 dB, rather than sampling uniformly from ranges such as –5 to 5 or –10 to 10 dB. This suggests that 0 dB provides a balanced contrast between machine and noise, enabling robust representations, whereas sampling from wider ranges dilutes this contrast and weakens performance.

\input{experiments/result_dataset_comparison}
Table~\ref{tab:dataset_comparison} compares the effect of different pre-training datasets. Pre-training with MIMII-DG \cite{dohi_mimii_2022} and ToyADMOS+ \cite{haradatoyadmos2+} clean machine sounds yields better overall performance than using AS-20K (70.5 vs. 61.5), suggesting that machine sounds play an important role in pre-training audio encoders for ASD. The best performance is achieved when training includes the same domain data (both machine and noise) as in the evaluation set (71.8), indicating that domain-matched pre-training provides a strong advantage by reducing the domain gap between training and testing.

\label{sec:majhead}
\begin{figure}
    \begin{minipage}[b]{1.0\linewidth}
      \centering
      \centerline{\includegraphics[width=8.5cm]{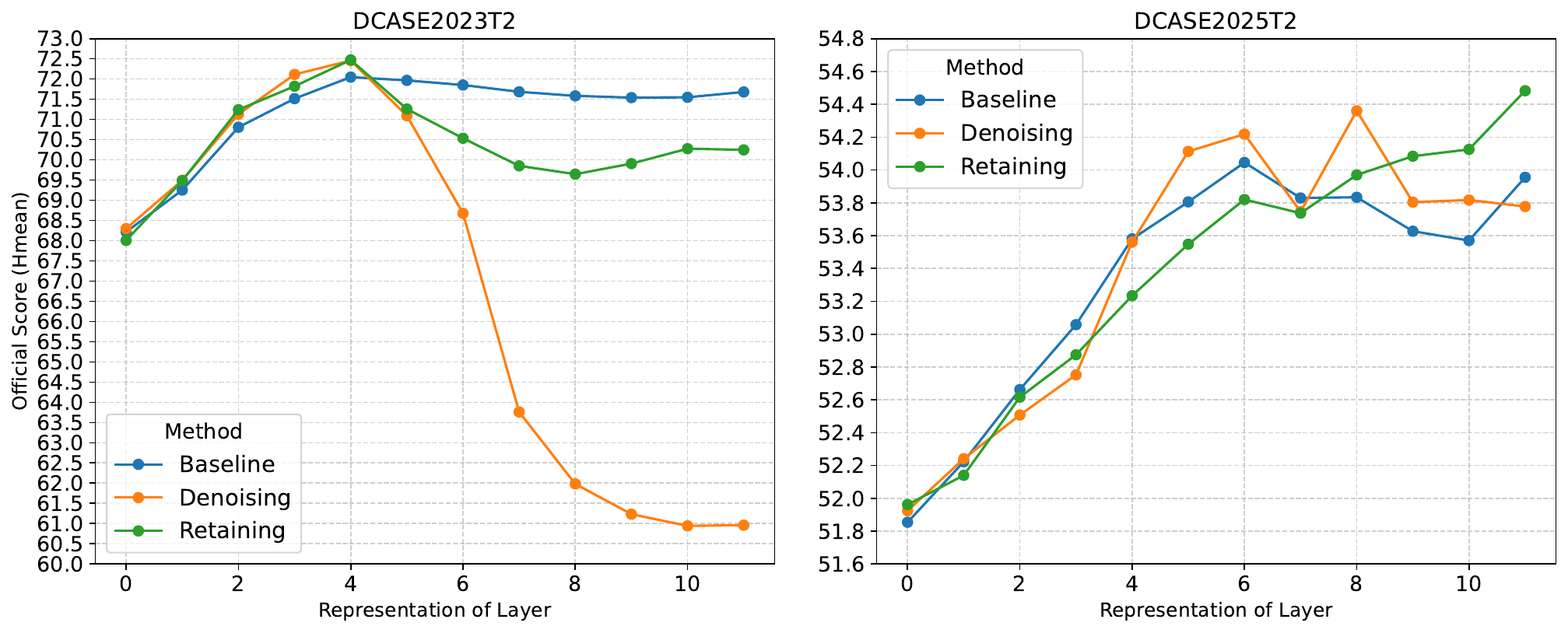}}
      \vspace{-3mm}
    \end{minipage}
    \caption{Comparison of \textit{GenRep} \cite{genrep} performance using the BEATs audio encoder \cite{pmlr-v202-chen23ag} (blue), the denoising baseline (orange), and the proposed retention method (green) on the DCASE2023T2 evaluation set (left) \cite{dohi_description_2023} and the DCASE2025T2 evaluation set (right) \cite{Nishida_arXiv2025_01}.}
    \label{fig:layereffect}
\vspace{-4mm}  
\end{figure}

Figure~\ref{fig:layereffect} shows analysis on the public benchmark evaluation sets \cite{dohi_description_2023, Nishida_arXiv2025_01}, comparing baselines with our method under the training-free condition. Two key findings emerge: (i) our method preserves performance in later layers, whereas the denoising baseline suffers degradation. This highlights that representation alignment is a more effective objective since it retains information useful for downstream evaluation, consistent with \cite{bolya2025perceptionencoderbestvisual} where aligning the last layer to the teacher’s best representation improved task performance. (ii) the best-performing layers for our method (layers 4 and 12) matched or exceeded those of the baseline (layers 4 and 8), demonstrating consistent advantages across machine types. Gains on these public benchmarks are modest, likely due to diverse noise and SNR conditions, but they still indicate that representation alignment enhances robustness. Broader exploration with larger and more varied training data may further amplify these improvements.

%% file: experiments/result_feature_alignment.tex

\begin{table}[!htbp]
\centering
\caption{Analysis of feature alignment. Performances are reported using the official score under the \textit{GenRep} \cite{genrep} framework, without applying score normalization between source and target domains.}
\label{tab:feature_alignment}
\setlength{\tabcolsep}{5pt}
\resizebox{\columnwidth}{!}{%
\begin{tabular}{lccccccc}
\toprule
& \multicolumn{6}{c}{Mismatch at 0 dB} \\
\cmidrule{2-7}
Method &Bandsaw &BandSealer &Grinder &Polisher &Screwfeeder &Shaker &Hmean \\\midrule
Wave mixture (BEATs iter3 \cite{pmlr-v202-chen23ag, genrep}) &57.6 &47.0 &82.6 &57.8 &72.2 &57.6 &60.4 \\
Embedding mixture (oracle) &70.6 &63.1 &69.2 &75.7 &72.4 &99.7 &73.6 \\
\textit{Tagging Loss} &50.6 &50.8 &90.2 &56.6 &64.6 &82.3 &62.6 \\
\textit{Mixture Loss} &57.0 &48.6 &89.2 &58.7 &66.9 &82.2 &64.2 \\
\textit{Tagging Loss} + \textit{Mixture Loss} &57.9 &49.5 &87.3 &59.0 &67.2 &74.5 &63.7 \\
\bottomrule
\end{tabular}
}
\end{table}

%% file: experiments/result_snr_abla.tex

\begin{table}[!htp]\centering
\caption{Comparison on different target SNRs for Eq.~(3) using \textit{Tagging Loss.}}\label{tab:result_snr_ablation}
\setlength{\tabcolsep}{5pt}
\resizebox{0.83\columnwidth}{!}{%
\begin{tabular}{lrrrrrrrrr}\toprule
&\multicolumn{7}{c}{Mismatch} & \\\cmidrule{2-8}
Target SNR mixture &-10 &-5 &0 &5 &10 &20 &30 &Hmean \\\midrule
Baseline \cite{pmlr-v202-chen23ag, genrep} &47.7 &50.9 &61.5 &73.8 &83.5 &91.5 &94.4 &67.4 \\
SNR at $10$ dB &51.1 &56.3 &\textbf{66.9} &74.8 &83.5 &92.4 &95.4 &70.8 \\
SNR at $5$ dB &51.0 &51.3 &60.1 &70.7 &81.7 &94.1 &96.3 &67.9 \\
SNR at $0$ dB &\textbf{53.3} &\textbf{57.8} &65.6 &\textbf{77.9} &\textbf{87.0} &\textbf{95.4} &\textbf{97.2} &\textbf{72.6} \\
SNR at $-5$ dB &48.3 &47.9 &59.7 &74.3 &85.3 &95.0 &96.7 &67.1 \\
SNR at $-10$ dB &48.0 &51.4 &57.6 &74.6 &86.7 &94.7 &96.4 &67.6 \\
SNR at $\pm 5$ dB &49.1 &55.9 &62.8 &71.2 &81.0 &91.1 &94.6 &68.5 \\
SNR at $\pm 10$ dB &44.7 &42.9 &55.3 &70.1 &83.1 &92.5 &94.7 &62.8 \\
\bottomrule
\end{tabular}
}
\vspace{-2mm}
\end{table}

%% file: experiments/result_dataset_comparison.tex
\begin{table}[!t]\centering
\caption{Comparison of different pre-training data using \textit{Mixture Loss}.
}\label{tab:dataset_comparison}
\setlength{\tabcolsep}{5pt}
\resizebox{0.9\columnwidth}{!}{%
\begin{tabular}{lccccccccc}\toprule
&\multicolumn{7}{c}{Mismatch} \\\cmidrule{2-8}
Pre-training data &-10 &-5 &0 &5 &10 &20 &30 &Hmean \\\midrule
AS-2M (Baseline \cite{pmlr-v202-chen23ag}) &47.7 &50.9 &61.5 &73.8 &83.5 &91.5 &94.4 &67.4 \\
\hspace{0.5em}+ AS-20K &45.4 &46.5 &52.7 &63.6 &72.6 &88.1 &94.8 &61.5 \\
\hspace{0.5em}+ Pre-training machine data  &50.7 &56.6 &64.6 &74.6 &84.4 &92.9 &95.7 &70.5 \\
\midrule
\hspace{0.5em}+ Eval-reference data &51.1 &59.8 &67.9 &75.9 &84.5 &91.5 &94.6 &71.8 \\
\bottomrule
\end{tabular}
}
\vspace{-2mm}
\end{table}